%% file: paper.tex
\documentclass[aps,prl,reprint,superscriptaddress, showpacs]{revtex4-1}
\usepackage{amsmath}
\usepackage{amssymb}
\usepackage{bm}
\usepackage{graphicx}
\usepackage{pdfpages}

\newcommand{\mvec}[1]{\ensuremath{\mathbf{#1}}}

\newcommand{\Heff}{\ensuremath{\mvec{H}_\mathrm{eff}}}

\newcommand{\mean}[1]{\left< {#1} \right>}

\begin{document}
\title{Magnon-Driven Domain-Wall Motion with the Dzyaloshinskii-Moriya Interaction}

\author{Weiwei Wang}
\affiliation{Engineering and the Environment, University of Southampton,  SO17 1BJ, Southampton, United Kingdom}
\author{Maximilian Albert}
\affiliation{Engineering and the Environment, University of Southampton,  SO17 1BJ, Southampton, United Kingdom}
\author{Marijan Beg}
\affiliation{Engineering and the Environment, University of Southampton,  SO17 1BJ, Southampton, United Kingdom}
\author{Marc-Antonio Bisotti}
\affiliation{Engineering and the Environment, University of Southampton,  SO17 1BJ, Southampton, United Kingdom}
\author{Dmitri Chernyshenko}
\affiliation{Engineering and the Environment, University of Southampton,  SO17 1BJ, Southampton, United Kingdom}
\author{David Cort\'es-Ortu\~no}
\affiliation{Engineering and the Environment, University of Southampton,  SO17 1BJ, Southampton, United Kingdom}
\author{Ian Hawke}
\affiliation{Mathematical Sciences, University of Southampton,  SO17 1BJ, Southampton, United Kingdom}
\author{Hans Fangohr}
\email{fangohr@soton.ac.uk}
\affiliation{Engineering and the Environment, University of Southampton,  SO17 1BJ, Southampton, United Kingdom}

\begin{abstract}
We study domain wall (DW) motion induced by spin waves (magnons) in the presence of Dzyaloshinskii-Moriya interaction (DMI).
The DMI exerts a torque on the DW when spin waves pass through the DW, and
this torque represents a linear momentum exchange between the spin wave and the DW.
Unlike angular momentum exchange between the DW and spin waves,
linear momentum exchange leads to a rotation of the DW plane
rather than a linear motion. In the presence of an effective easy
plane anisotropy, this DMI induced linear momentum transfer mechanism is significantly more
efficient than angular momentum transfer in moving the DW.
\end{abstract}


\maketitle

The manipulation of domain wall (DW) motion has been extensively studied in the past few years
due to potential applications in logic devices and data storage technology \cite{Allwood2005a, Hertel2004, Wieser2004, Parkin2008, Hu2013}.
A DW can be driven by an applied field \cite{Schryer1974},
microwaves \cite{Yan2009}, spin transfer torque \cite{Zhang2004} and
spin waves (magnons) \cite{Hinzke2011,Han2009a,Yan2011}.
Spin waves can drive the DW effectively since they carry magnonic spin current.
In general, when the spin waves travel through the DW, the DW acquires a negative velocity -- relative to the propagation
direction of the spin waves -- due to
conservation of angular momentum \cite{Yan2011}, although positive
velocities have been observed
in micromagnetic simulations at special frequencies \cite{Han2009a,Wang2012a, Wang2012b, Kim2012}.

Angular momentum conservation plays a crucial role in spin wave induced DW motion:
when the spin wave passes through the DW, the magnonic spin current
changes its sign, which generates a torque and the DW moves in order to absorb this torque. 
Magnons can be considered as particles with angular momentum~$\pm \hbar$ and linear momentum~$\hbar k$ \cite{Yan2011}. 
When the spin wave is reflected, linear momentum is transferred to the DW which results in DW motion \cite{Wang2012a,Yan2013}.
The difference between these two mechanisms is that the DW moves in opposite directions \cite{Yan2013, Janutka2013}.
In this Letter we demonstrate, by using micromagnetic simulations and a one-dimensional (1d) analytical DW model, 
that spin waves passing through a domain wall in the presence of Dzyaloshinskii-Moriya interaction (DMI) and 
an easy-plane anisotropy drive the domain wall very effectively. We attribute this to linear momentum transfer 
and show that this effect can be more efficient than the better known angular momentum transfer by an order of magnitude.

The DMI is an anti-symmetric interaction induced by spin-orbit coupling due
to broken inversion symmetry in lattices or at the interface of magnetic films \cite{Fert2013}.
The DMI can  lead to chiral magnetic orders such as skyrmions and spin spirals \cite{Zang2011, Fert2013, Moon2013,Rohart2013}.
In addition, the DMI has brought new phenomena for DW dynamics driven by fields \cite{Thiaville2012}
or charge currents \cite{Tretiakov2010}.
The DMI has been found both for magnetic interfaces \cite{Rohart2013} and bulk materials such as 
MnSi \cite{Muhlbauer2009} and FeGe \cite{Huang2012}. In this work we focus on bulk DMI with
micromagnetic energy density $\varepsilon_{\mathrm{dmi}}=D \mvec{m}\cdot (\nabla \times \mvec{m})$ where
$D$ is the DMI constant and $\mvec{m}$ is the normalized magnetization.

We consider a quasi-1d nanowire with exchange interaction, DMI and two effective anisotropies. 
 One anisotropy $K$ is the uniaxial anisotropy along
the $x$-axis, and the other effective $K_\perp$ is an easy $xy$-plane anisotropy. 
The combined anisotropies can be considered as a model of overall effect including the demagnetization field, 
surface or magnetoelastic anisotropy \cite{Huang2012, Porter2014}.
The total free energy for the wire along the $x$-axis is
\begin{equation}\label{eq_energy}
E=S\int \left[A (\nabla \mvec{m})^2 - K m_x^2 + K_\perp  m_z^2 + \varepsilon_{\mathrm{dmi}} \right] \mathrm{d}x,
\end{equation}
where $S$ is the cross-sectional area of the wire and $A$ is the exchange constant.

The dynamics of the magnetization $\mvec{m}$ is governed by the Landau-Lifshitz-Gilbert (LLG) equation
\begin{equation}\label{eq_LLG}
\frac{\partial \mvec{m}}{\partial t}=
-\gamma \mvec{m} \times \mvec{H}_{\mathrm{eff}}
+\alpha\mvec{m} \times \frac{\partial \mvec{m}}{\partial t},
\end{equation}
where $\gamma\;(>0)$ is the gyromagnetic ratio and $\alpha$ is the Gilbert damping.
The effective field $\Heff$ is calculated as
the functional derivative $\Heff=-1/(\mu_0 M_s) \delta E/\delta \mvec{m} =2/(\mu_0 M_s) [A \nabla^2 \mvec{m} - D \nabla \times \mvec{m} 
+  K m_x \mvec{e}_{x}- K_\perp m_z  \mvec{e}_{z}]$ with $M_s$ the
saturation magnetization and $\mu_0$ the vacuum permeability.

\begin{figure}[tbhp]
\begin{center}
\includegraphics[scale=0.52]{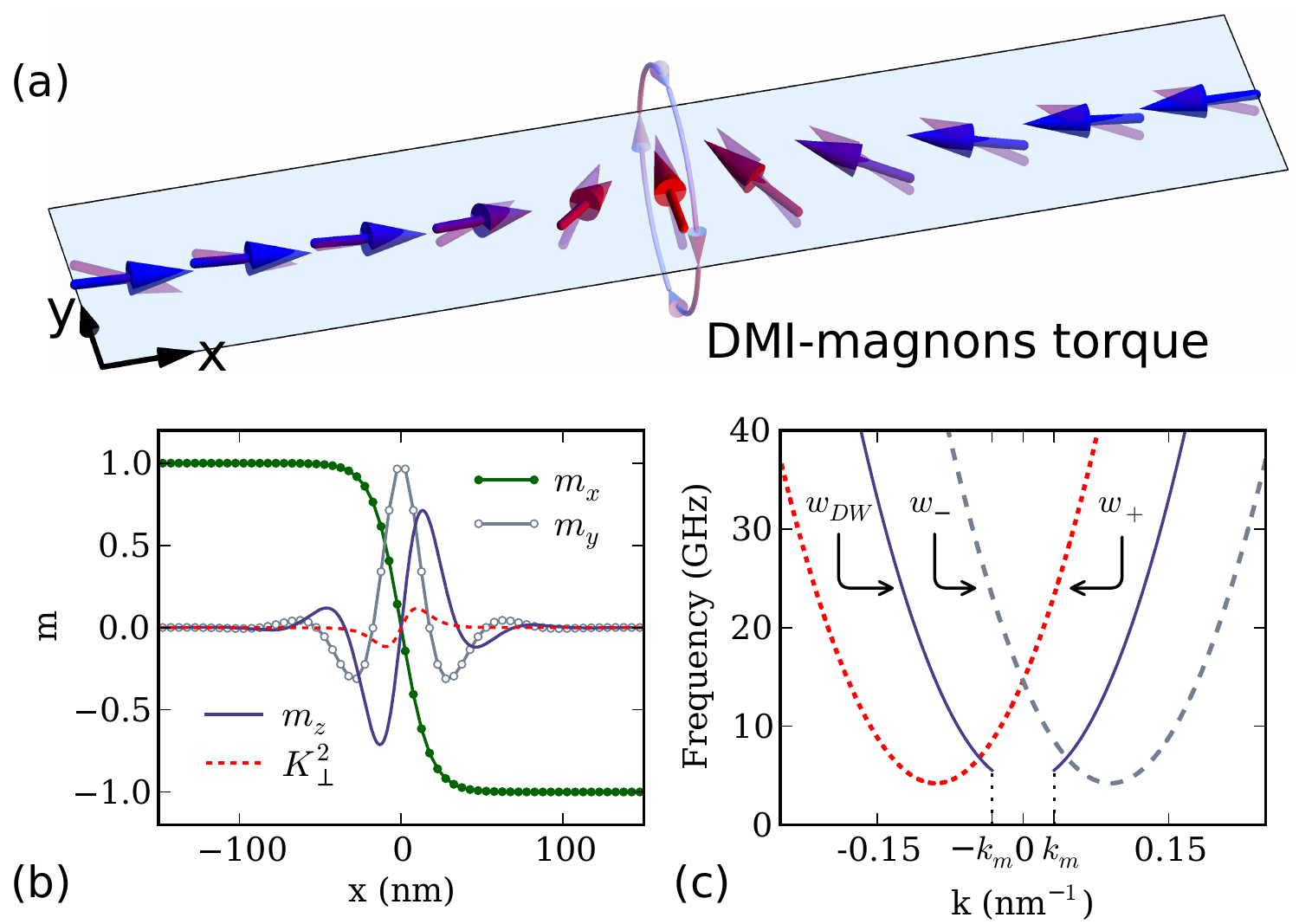}
\caption{ (a) Illustration of the head-to-head DW in the nanowire using red-blue opaque arrows.
The translucent purple arrows represent a spin wave excitation. The DMI exerts a torque to
change the DW tilt angle when spin waves pass through the DW.
(b) DW profile using Eq.\eqref{eq_dw} with parameters
$A = 8.78 \times 10^{-12}\,\mathrm{J/m}$, $K=1\times 10^{5}\, \mathrm{J/m^3}$,
$D=1.58\times 10^{-3} \mathrm{J/m^2}$, $K_\perp = 0$ and $\Phi=0$.
The red dashed line shows the simulation data for $m_z$ with $K_\perp^2=
6\times 10^5\, \mathrm{J/m^3}$: the easy-plane anisotropy
favours a reduced $m_z$.
(c) The dispersion relations inside and outside the DW.}
\label{fig_dw}
\end{center}
\end{figure}
The typical DW structures described by the energy\;(\ref{eq_energy}) for the case $D=0$ are head-to-head
and tail-to-tail DWs,  and the former is shown in Fig.~\ref{fig_dw}(a).  By using spherical
coordinates $\theta=\theta(x)$ and $\phi=\phi(x)$, the magnetization unit vector $\mvec{m}$
is expressed as $\mvec{m}=(\cos \theta, \sin \theta \cos \phi, \sin \theta \sin \phi)$,
and the total micromagnetic energy (\ref{eq_energy}) reads
\begin{align}\label{eq_energy2}
E= S \int \Bigl [A (\theta'^2+\sin^2 \theta \phi'^2)
- D \phi' \sin^2 \theta  \nonumber \\
+ K \sin^2 \theta (1+ \kappa \sin^2 \phi)
  \Bigr] \mathrm{d}x ,
\end{align}
where $\kappa=K_\perp/K$ and $'$ represents the derivative with respect to $x$.
In equilibrium state, the energy (\ref{eq_energy2}) must be minimal, and thus 
 we arrive at two coupled differential equations for $\theta$ and $\phi$ by using standard variational calculus,
\begin{gather}
\begin{split}\label{eq_diff}
2A \theta'' = \sin 2\theta (A \phi'^2 + K(1+ \kappa\sin^2\phi) -D \phi'),\\
\sin\theta (2A \phi'' - K_\perp \sin2\phi)= 2\cos\theta (D-2A\phi')\theta'.
\end{split}
\end{gather}
The corresponding boundary conditions are $\theta ' =0 $ and $ (\phi' - 1/\xi) \sin^2 \theta =0 $ for $x=\pm \infty $ 
(see Supplemental Material \cite{Supp}) where  $\xi=2A/D$ is the characteristic length \cite{Rohart2013}.
We are searching for the head-to-head DW solution, therefore the ansatz $\cos \theta = -\tanh[(x-x_0) /\Delta]$ is used, where
$\Delta$ is the DW width and $x_0$ is the DW center.
Initially, we consider the case of $\kappa=0$ (i.e. $K_\perp = 0$) which preserves the rotational symmetry.
We assume that $\phi$ is a linear function of space $x$, i.e., $\phi(x)=(x-x_0)/\xi+\Phi$ where $\Phi$ is
the DW tilt angle. Inserting it back to Eq.~\eqref{eq_diff} we obtain $\Delta = \sqrt{A/(K-A/\xi^2)}$.
 In the absence of DMI, the DW width reduces to $\Delta_0=\sqrt{A/K}$
which is the well known Bloch wall width.
Therefore the static one-dimensional head-to-head DW profile can be expressed as \cite{Tretiakov2010} 
\begin{gather}
\begin{split}\label{eq_dw}
m_x &= - \mathrm{tanh}(x/\Delta), \\
m_y &= \mathrm{sech}(x/\Delta) \cos(x/\xi+\Phi),\\
m_z &= \mathrm{sech}(x/\Delta) \sin(x/\xi+\Phi),
\end{split}
\end{gather}
where we have chosen $x_0=0$.
Fig.~\ref{fig_dw}(b) shows the DW profile using Eq.~\eqref{eq_dw} for $K_\perp = 0$ with lines,
and the red dashed line depicts the micromagnetic simulation result of $m_z$ for $K_\perp^2 = 6\times 10^5\, \mathrm{J/m^3}$.
The rotational symmetry breaks for $K_\perp>0$ and the $z$-component
of the magnetization $m_z$ is suppressed by the easy plane anisotropy.
The DW configuration (\ref{eq_dw}) is not stable if the DMI constant is
larger than the critical value $D_c=2\sqrt{AK}$ \cite{Tretiakov2010},
and the presence of $K_\perp > 0$ increases this threshold.

We assume that the spin wave can be described by a small fluctuation $u=u(x)$ and $v=v(x)$ around
$\mvec{m}_0$,
where $\mvec{m}_0=(\cos \theta_0, \sin \theta_0 \cos \phi_0, \sin \theta_0 \sin \phi_0)$ is
the static domain wall profile Eq.~\eqref{eq_dw},
\begin{equation}\label{eq_m}
\mvec{m} = \mvec{m}_0 + [u(x) \mvec{e}_\theta + v(x)  \mvec{e}_\phi ] e^{-i w t},
\end{equation}
where $\sqrt{u^2+v^2} \ll 1$, $\mvec{e}_\phi = (0,-\sin \phi_0,\cos \phi_0)$,
 $\mvec{e}_\theta = (- \sin \theta_0, \cos \theta_0 \cos\phi_0, \cos \theta_0 \sin\phi_0)$,
and $\omega$ is the spin wave frequency.
By following the treatment in Ref.~[\citenum{Yan2011}], we obtain for the $K_\perp=0$ case,
\begin{gather}
\begin{split}\label{eq_linearize}
A  v'' - \tilde{K} v \cos(2\theta_0) &= -i u \omega/\gamma_0, \\
A  u'' - \tilde{K} u \cos(2\theta_0) &=  i v \omega/\gamma_0,
\end{split}
\end{gather}
where we define $\tilde{K}=K-D^2/(4A)$ and $\gamma_0 = 2 \gamma/(\mu_0 M_s)$.
By introducing the complex variable $\psi = u-iv$, Eq.~\eqref{eq_linearize}
can be written as a time-independent Schr\"{o}dinger-type equation with reflectionless potential \cite{Braun1994,Lekner2007},
\begin{align}\label{eq_sch}
\hat{H} \psi(\zeta) = (1+q^2) \psi(\zeta),
\end{align}
where $\zeta=x/\Delta$ and the operator is $\hat{H}=-d^2/d\zeta^2 +1 -2\,\mathrm{sech}^2(\zeta)$.
The eigenvalues $1+q^2=\omega/(\gamma_0 \tilde{K})$  define the spin wave dispersion relation inside the DW,
which is plotted in  Fig.~\ref{fig_dw}(c) (darkslateblue line) with wavevector $k=q/\Delta$.
The above discussion is only valid for wavelengths smaller than the domain wall size,
which corresponds to wave vectors greater than $k_m \sim 1/(2\Delta)$.
The propagating wave excitations can be
expressed as $\psi(\zeta,t)=\rho_k e^{i \Omega} (\tanh(\zeta)- iq)$ where $\Omega=\zeta q-\omega t$ represents the sine or cosine type waves
 and $\rho_k$ the wavevector dependent spin wave amplitude \cite{Tveten2014}. The reflectionless property for spin waves
holds even in the presence of the easy plane anisotropy \cite{Yan2012}.
Interestingly, the dispersion relation inside the DW is symmetric in
the reduced wavevector $q$ even though the wall is twisted by the DMI.
However, due to the exponential decay of the DW profile when moving away from the DW centre, the magnetization is
uniform in the domains and the dispersion relations become asymmetric
outside the DW \cite{Zakeri2010,Moon2013},
\begin{align}\label{eq_dispersion}
\omega_{\pm} = \gamma_0(K + A k^2 \pm D k).
\end{align}
Fig.~\ref{fig_dw}(c) shows the asymmetric dispersion relations outside the DW.
The dispersion relation  (\ref{eq_dispersion}) also suggests that the wavevector
changes by $D/A$ when the spin wave passes through the DW if the frequency of the spin wave remains the same.
The spin wave becomes elliptical rather than circular if $K_\perp>0$ and the corresponding dispersion relation
outside the DW becomes $\omega_{\pm} = \gamma_0[\sqrt{(K + A
  k^2)(K+K_\perp + A k^2)} \pm D k]$ \cite{Moon2013}.

To study the DW dynamics, micromagnetic simulations have been performed using a 1d mesh with length 2000 nm and cell size 1 nm.
We make use of the parameters of FeGe \cite{Beg2014}: the exchange constant $A = 8.78 \times 10^{-12}\,\mathrm{J/m}$,
the DMI constant $|D|=1.58\times 10^{-3} \mathrm{J/m^2}$, 
the saturation magnetization $M_s = 3.84\times 10^5\,\mathrm{A/m}$ and the damping coefficient  $\alpha=0.01$.
We set the easy axis anisotropy $K=1\times 10^5\, \mathrm{J/m^3}$ and treat $K_\perp$ as an adjustable parameter since
both anisotropies depend on the sample shape, strain and surface effects \cite{Majkrzak2012}.
The spin waves are excited locally in the region $-1000 \leq x \leq -998\,\textrm{nm}$ by a linearly polarized 
field $\mvec{h}(t)=h_0 \sin(2 \pi f  t) \mvec{e}_y$ with $h_0=1\times 10^5\,\mathrm{A/m}$. 
The initial domain wall is located at $x_0=0$, and to prevent spin wave reflection the damping 
coefficient is increased linearly \cite{Han2009a} from 0.01 to 0.5  in the region $800 \leq x \leq 1000\,\textrm{nm}$.

\begin{figure}[tbhp]
\begin{center}
\includegraphics[scale=0.72]{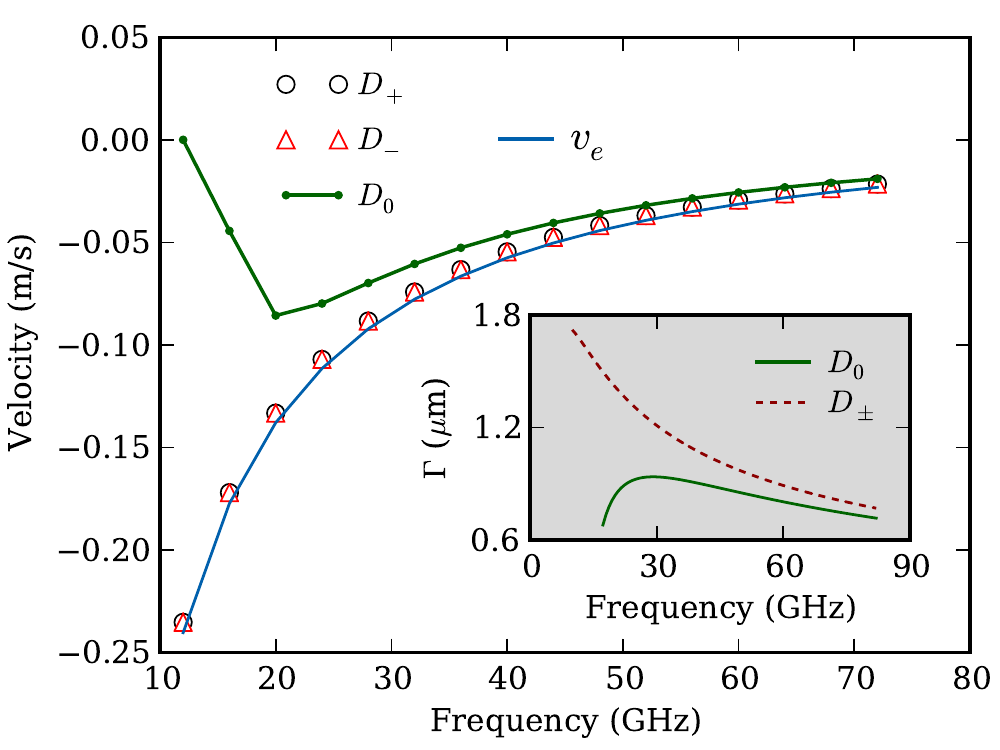}
\caption{Simulation results of the DW velocity as a function of spin wave frequency with different DMI constants for the case
of $K_\perp=0$. The DMI parameters are $D_0=0$ and $D_\pm=\pm 1.58\times 10^{-3} \mathrm{J/m^2}$.
The $v_e$ curve is calculated by $v_e=-\frac{\rho^2}{2} V_g$  \cite{Yan2011} where $\rho$ at $x=0$ is extracted from the simulation.
Inset: Plot of spin wave amplitude decaying characteristic length $\Gamma$ versus frequency.}
\label{fig_v}
\end{center}
\end{figure}
The spin wave traveling in the $+x$ direction induces DW motion.
Fig.~\ref{fig_v} shows the DW velocity as a function of frequency with
different DMI constants for $K_\perp=0$. The DW velocity is negative, which is explained by conservation of
angular momentum, and the DW velocity is $v_e=-\frac{\rho^2}{2} V_g$  \cite{Yan2011}, where $V_g=\frac{\partial \omega_k}{\partial k} $
 is the spin wave group velocity and $\rho$ is the spin-wave amplitude.
For a circular spin wave, i.e. for $K_\perp=0$, by using the dispersion relation
inside the DW or Eq.~\eqref{eq_dispersion} we have $V_g=2\gamma_0 A k = 2\sqrt{\gamma_0 A (\omega - \gamma_0 \tilde{K})}$. 
In the absence of DMI, the DW velocity is zero if the frequency is less than the cut-off frequency $f_\mathrm{cut}=\gamma_0 K \approx 14.5$ GHz,
which is reduced to $\gamma_0 \tilde{K} \approx 4.2$ GHz by DMI. 
The magnitude of the DW velocity first increases, and then decreases as the frequency of the spin wave increases.
The reason for this is that the spin wave amplitude decays exponentially as the spin wave propagates.
To quantify this, we assume the magnetization has the form  
  $\mvec{m}=\pm \mvec{e}_x + \bm{\rho_0} e^{i (k x - \omega t)}e^{-x/\Gamma}$ with $|\bm{\rho_0}| \ll 1$ \cite{Seo2009, Moon2013}, 
and obtain $\Gamma_{\pm} = 2/(\alpha \omega) [\gamma_0 A k  \pm D(\omega \mp D \gamma_0 k)/(K_\perp +2K + 2A k^2)]$,
which is plotted in the inset of Fig.~\ref{fig_v} with $K_\perp=0$ and shows that the spin wave amplitude decaying is reduced by
the existence of DMI. The predicted DW velocity $v_e$ is plotted in Fig.~\ref{fig_v} as well, which fits the simulation results very well.
\begin{figure}[tbhp]
\begin{center}
\includegraphics[scale=0.72]{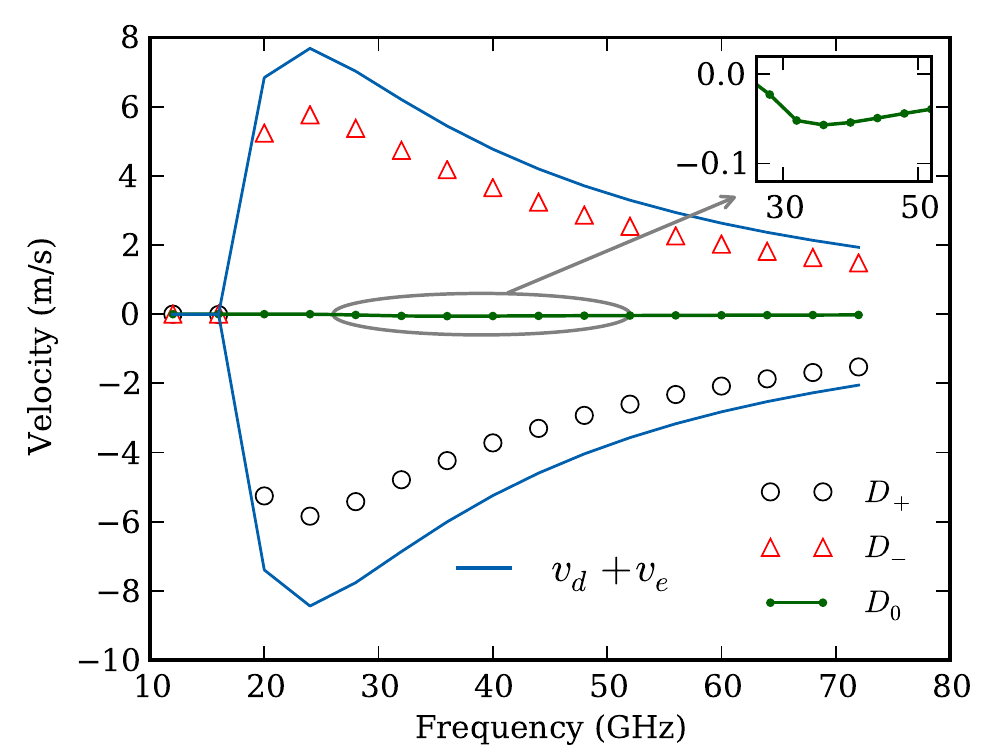}
\caption{The DW velocity as a function of the spin wave frequency with $K_\perp= 2\times 10^5\, \mathrm{J/m^3}$. 
The DMI constants employed in the simulation are $D_0=0$ and $D_\pm=\pm 1.58\times 10^{-3} \mathrm{J/m^2}$.}
\label{fig_v_kp}
\end{center}
\end{figure}

We now repeat the study for Fig.~\ref{fig_v} above with $K_\perp>0$ and where the spin waves are elliptical.
Fig.~\ref{fig_v_kp} shows the DW velocity as a function of spin wave frequency for $K_\perp= 2\times 10^5\, \mathrm{J/m^3}$,
and the corresponding DW displacements are shown in Fig.S1 (see Supplemental
Material \cite{Supp}). 
As in the $K_\perp=0$ case, we find no spin wave reflection,
and the DW velocity is negative if the DMI constant $D$ is $>0$.
Similar to the $K_\perp=0$ case, the DW velocity is zero when frequency $f<f_\mathrm{cut} \approx 16$ GHz,
and the DW velocity first increases, and then decreases with the frequency. 
However, the magnitudes are significantly larger, and for the $D<0$
case the DW velocity is positive.

To understand this novel DMI induced linear momentum transfer phenomenon, we recall the dispersion relation (\ref{eq_dispersion}) outside the DW and assume the
wavevector of a magnon before and after passing through the DW to be $k_1$ and $k_2$, respectively.
When spin waves travel though the DW, they jump from branch $\omega_+$ to $\omega_-$ in the dispersion relation, as depicted in Fig.~\ref{fig_dw}(c) 
or Fig.~S2(a) in \cite{Supp}. By assuming the frequency keeps the same, the change in wavevector $\delta k = k_2-k_1$ can be calculated.
 We show in Fig.~S3(a) \cite{Supp} that the frequency does not change significantly for our system.
The change in wavevector $\delta k$ leads to a momentum change $\delta
p= \hbar \delta k$ for each magnon.

The excited magnon density is $n=\rho^2 M_s/(2\hbar \gamma)$ \cite{Wang2012a}
and for elliptical spin waves we choose $\rho^2=u_0 v_0$ where $u_0$, $v_0$ are fluctuation amplitudes in $\mvec{e}_\theta$ and $\mvec{e}_\phi$.
The linear momentum of a DW is
$P_\mathrm{DW} = M_s/\gamma \int\phi \sin\theta (\partial \theta/\partial x)\mathrm{d} x
= 2 \Phi M_s/\gamma$ \cite{Kosevich1990} and
conservation of linear momentum \cite{Yan2013} gives $d P_\mathrm{DW} /dt=- dP_\mathrm{magnons}/dt = - n V_g \delta p$, i.e.,
$\dot{\Phi}= - \frac{1}{4} \rho^2 V_g \delta k$.
To describe the domain wall motion, we introduce an effective field along the $x$ direction by using the spherical form of the LLG equation,
\begin{equation}\label{eq_effective_field}
H_x=\dot{\Phi}/\gamma=-\frac{1}{4} \rho^2 \delta k V_g /\gamma.
\end{equation}
For circular spin waves $\delta k = D/A$, and thus the corresponding effective field is $H_x^0=-\frac{\rho^2}{2} D k \gamma_0/\gamma$.
In the $\kappa>0$ case (i.e.\ for $K_\perp>0$), the spin wave is elliptical and  $\delta k$ is a function of the frequency
(see  Fig.~S2  in Supplemental Material \cite{Supp}).
The presence of a non-zero $K_\perp$ suppresses the wavevector change, especially for low frequency spin waves.
The DW velocity $v_d$  induced by this effective field $H_x$
 in the presence of damping can be obtained using the rigid DW model \cite{BurkardHillebrands2006},
\begin{equation}\label{eq_v}
v_d =\frac{\gamma \Delta H_x}{\alpha} \Big/ \sqrt{1+\frac{\kappa}{2}\left(1-\sqrt{1-h^2} \right)},
\end{equation}
where $h=H_x/(\alpha H_{K_\perp})$ and $H_{K_\perp}=2K_\perp/(\mu_0 M_s)$. 
The total velocity is the sum of the established $v_d$ and $v_e$, which correspond to the linear and angular momentum conservation, respectively.

To estimate the total velocity $v_e + v_d$,  we have extracted the spin wave amplitude
 $\rho$ at $x=0$ (the initial position of the domain wall) from the simulation and the constant DW width $\Delta_0$ is used. 
This total velocity is shown as lines in Fig.~\ref{fig_v_kp} and shows a good
agreement with the simulation results shown as circle and triangle symbols. 
The maximum DW velocity is around $f=24$ GHz, which originates from the combined dependencies of $V_g$, $\Gamma$ and $\delta k$. 
Fig.~S2(b) shows that $\delta k$ does not change significantly as the frequency increases.
The DW can rotate freely if $K_\perp=0$  and the DW velocity induced by the field $H_x$ is
$v_0=\alpha \Delta \gamma_0 H_x/(1+\alpha^2)$. We can establish that
$v_0 \sim 10^{-4}$ m/s,  which could explain why the linear momentum exchange is not significant for the DW motion shown in Fig.~\ref{fig_v}.

\begin{figure}[tbhp]
\begin{center}
\includegraphics[scale=0.56]{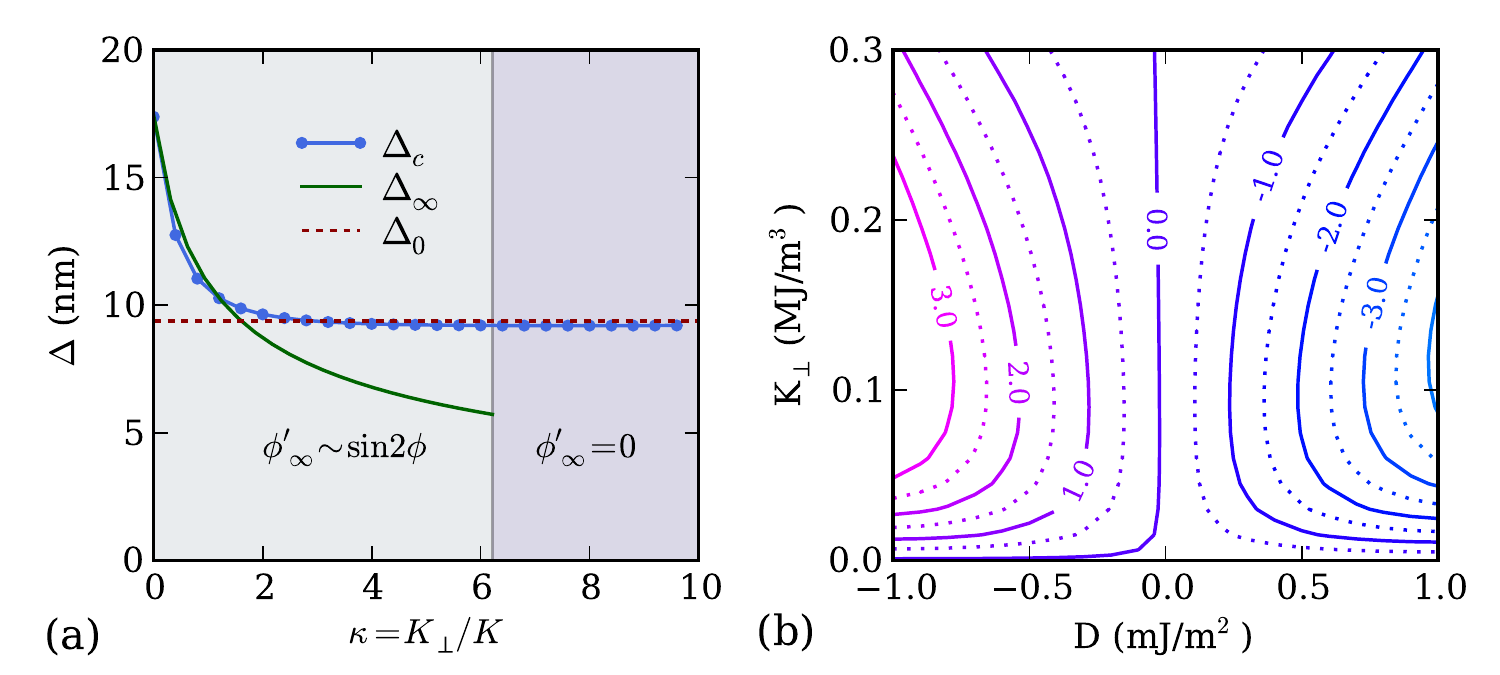}
\caption{(a) Plot of two types of domain walls, $\Delta_c$ is obtained by fitting the simulation data with $\cos\theta =- \tanh (x/\Delta_c)$.
(b) The contour plot of the simulated DW velocity (in m/s) for different $K_\perp$ and DMI constants, where the frequency of the external ac field is fixed 
at $f=30$ GHz.}
\label{fig_contour}
\end{center}
\end{figure}

The domain wall width $\Delta$ is not a constant for the $K_\perp>0$ case, and the corresponding DW profiles are described by Eq.~\eqref{eq_diff}.
Using the asymptotic behavior of Eq.~\eqref{eq_diff} (see Supplemental Material \cite{Supp}) we can identify two types
of domain walls when $K_\perp>0$ and $D \neq 0$. The first type of DW is $\phi'_\infty=0$ which corresponds to small
$|D|$ case with $\kappa>0$,  as shown in Fig.~\ref{fig_contour}(a).
The second type is $\phi'_\infty \sim \sin 2\phi$ where $\phi$ is a monotonic function. 
In this scenario, the DW width $\Delta_\infty$ for $x \rightarrow \infty$ is given by 
$1/\Delta_\infty^2=(1+\kappa/2+\sqrt{1+\kappa})K/(2A)-1/\xi^2$. From Fig.~\ref{fig_contour}(a) we can find that
$\Delta_\infty$ is a good approximation if $\kappa<2$.
The critical $\kappa_c$ can be obtained by solving the equation $AK \kappa_c^2=2D^2(1+\kappa_c/2+\sqrt{1+\kappa_c})$,
which gives $\kappa_c \approx 6.2$.  The simulation results also show that for $\kappa \gg 1$ the DW width $\Delta_c$ 
is close to $\Delta_0$.

So far the effective field is introduced by linear momentum conservation. In the following section we cross-check
this using the LLG equation.
The LLG equation (\ref{eq_LLG}) with zero damping is rewritten to describe the spin conservation law \cite{Tatara2008},
\begin{equation}\label{eq_llg2}
\frac{\partial \mvec{m} }{\partial t}+
\frac{\partial \mvec{j}_e}{\partial x} = \bm{\tau}_a + \bm{\tau}_d
\end{equation}
where $\mvec{j}_e =\gamma_0 A \mvec{m} \times \partial_x \mvec{m}$ is the exchange spin current associated with localized spin.
The spin source or sink
$\bm{\tau}_a= -\gamma_0 \mvec{m} \times [K m_x \mvec{e}_{x}- K_\perp m_z  \mvec{e}_{z}]$
and $\bm{\tau}_d= \gamma_0 D \mvec{m} \times ( \nabla \times \mvec{m})$  come from the anisotropy and DMI, respectively.
The average DW velocity can be computed through $v=  \frac{1}{2} \int \left<\frac{\partial m_x}{\partial t} \right>  \mathrm{d}x $ where $\mean{f(t)}$ represents the temporal average for a periodic function $f(t)$. 
To compute this average we keep the terms up to the square of the spin waves amplitude and ignore the higher-order ones. 
By integrating over space for the $x$-component of the spin current $\mvec{j}_e$, the velocity
$v_e$ can be recovered.

By using the  DW profile (\ref{eq_dw}) it is found that the overall contributions of the $x$-component torques $\bm{\tau}_a$ and $\bm{\tau}_d$ are zero,
i.e., $\int {\left< \bm{\tau}_a^x \right>} dx = \int {\left< \bm{\tau}_d^x \right>} dx = 0$.
However, the contribution of the $z$-component
of the DMI torque is nonzero, i.e.,
$\int {\left< \bm{\tau}_d^z \right>} dx = - \int \frac{\rho^2}{2} \gamma_0 D k  \, m_y   dx$,
which represents an additional torque rotating the DW plane. By introducing an effective field $H_x^0$ in
the $x$ direction such that the total torque on the DW equals the torque $\mvec{\tau}_d^z$, we obtain
$H_x^0  = \int {\left< \mvec{\tau}_d^z/\gamma \right>} dx \Big / \int m_y dx = -  \frac{\rho^2}{2} D k {\gamma_0}/{\gamma} $,
which is in exact agreement with the analysis above.

Figure~\ref{fig_contour}(b) shows a contour plot of the DW velocity as a function of $K_\perp$
and DMI constant~$D$. The figure is approximately symmetric in the DMI constant, with a biased velocity
originating from the angular momentum exchange between the spin wave and the DW.
The DW velocity is always negative if $D>0$. There exist some optimal
areas in which the DW has the highest velocity, and this area depends on the frequency of the spin wave. 

For a 2d magnetic sample, the magnetization at the edges is tilted due
to the DMI, and the domain wall velocity is slightly reduced compared to the
1d model used above (see Fig.~S5 in the Supplemental Material \cite{Supp}). 

In conclusion, we have studied DMI induced linear momentum transfer DW
motion. We find that the DMI exerts an extra torque which rotates the
DW plane when the spin wave passes through the DW, and that the
effective easy plane anisotropy supresses the rotation and leads to a
fast DW motion.  
The effect of the linear momentum is equivalent to an effective field and the direction of the field depends on the sign of the DMI constant and the DW profile.
This linear momentum exchange between spin waves and DW exists in addition
to the angular momentum exchange when magnons pass through the DW, and
is more efficient in moving the domain wall.

We acknowledge financial support from EPSRC's DTC grant EP/G03690X/1. 
W.W thanks the China Scholarship Council for financial assistance.

\bibliographystyle{apsrev4-1}
\input{paper.bbl}
\clearpage


\section*{Supplementary material}

## THE BOUNDARY CONDITIONS

In the equilibrium state, the total micromagnetic energy $E = S \int \mathcal{H} \, dx$ must be minimal, i.e., the Lagrangian density $\mathcal{L} = \mathcal{H}$ satisfies the Euler-Lagrange equation,

$$\frac{\partial L}{\partial f} - \frac{d}{dx}\frac{\partial L}{\partial f'} = 0, \tag{S1}$$

where $S$ is the cross-sectional area of the wire, $f = \theta, \phi$ and $\mathcal{H} = A(\theta'^2 + \sin^2\theta \phi'^2) - D\phi' \sin^2\theta + K\sin^2\theta(1 + \kappa \sin^2\phi)$. In the derivation of (S1), some extra surface terms appear in the integration by parts step as $f$ can vary at the boundary. Those terms vanish only if the following boundary conditions are imposed,

$$\frac{\partial L}{\partial f'} = 0 \quad \text{at the boundary.} \tag{S2}$$

Therefore, the boundary conditions are $\theta' = 0$ and $(2A\phi' - D)\sin^2\theta = 0$.

## THE DISPLACEMENT OF THE DOMAIN WALL

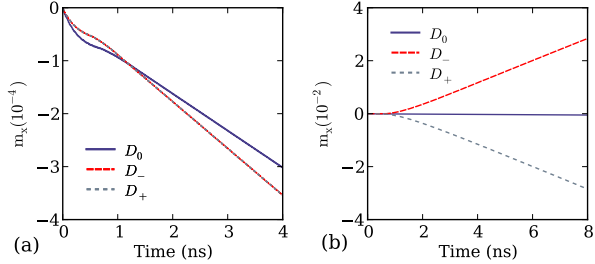

FIG. S1. The domain wall displacement for (a) $K_\perp = 0$ at frequency $f = 28$ GHz, and (b) $K_\perp = 2 \times 10^5$ J/m$^3$ at frequency $f = 36$ GHz. The DMI constants employed in the simulation are $D_0 = 0$ and $D_\pm = \pm 1.58 \times 10^{-3}$ J/m$^2$.

Fig. S1 shows the domain wall displacement with and without easy plane anisotropy for various DMI constants. As shown in Fig. S1(a), it is found that both negative and positive DMI constants lead to identical domain wall motion when $K_\perp = 0$. However, from Fig. S1(b) we can find that for the scenario of $K_\perp = 2 \times 10^5$ J/m$^3$, the sign of DMI constant influences the domain wall motion significantly. The domain wall velocity is obtained by fitting the displacement of the domain wall ($m_x$ as a function of time) and thus is calculated by $v = \frac{L}{2}\frac{dm_x}{dt}$ where $L$ is the length of the wire.

## THE DISPERSION RELATION AND THE CHANGES OF WAVEVECTOR

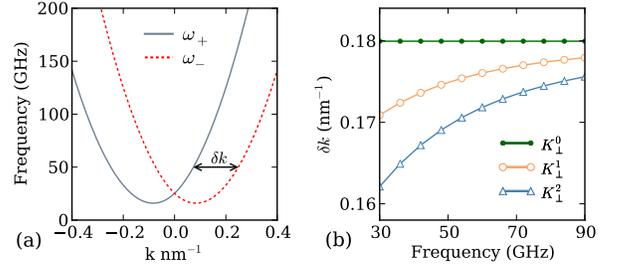

FIG. S2. (a) The dispersion relations with parameters $K_\perp = 2 \times 10^5$ J/m$^3$, $K = 1 \times 10^5$ J/m$^3$, and $D_\pm = \pm 1.58 \times 10^{-3}$ J/m$^2$. (b) Plot of $\delta k$ as a function of the frequency for $K_\perp^0 = 0$, $K_\perp^1 = 2 \times 10^5$ J/m$^3$ and $K_\perp^2 = 3 \times 10^5$ J/m$^3$ with $D = 1.58 \times 10^{-3}$ J/m$^2$.

Fig. S2(a) shows the dispersion relations for $K_\perp = 2 \times 10^5$ J/m$^3$ with positive and negative DMI constants. Initially, the spin waves follow the $\omega_+$ branch and after spin waves pass through the domain wall, the dispersion relation follows the $\omega_-$ branch, and thus the wavevector has been changed by $\delta k$. For the $K_\perp > 0$ case, $\delta k$ is a function of frequency, as shown in Fig. S2(b). The wavevector difference $\delta k$ can be computed from the dispersion relation

$$\omega_\pm = \gamma_0[\sqrt{(K + Ak^2)(K + K_\perp + Ak^2)} \pm Dk], \tag{S3}$$

by assuming the frequencies remain the same (i.e., $\omega_- = \omega_+$). For $K_\perp = 0$ case it is found that $\delta k = D/A$.

## THE CHANGES OF FREQUENCY

In the main article we have assumed that the frequency of spin waves does not change after the spin waves pass

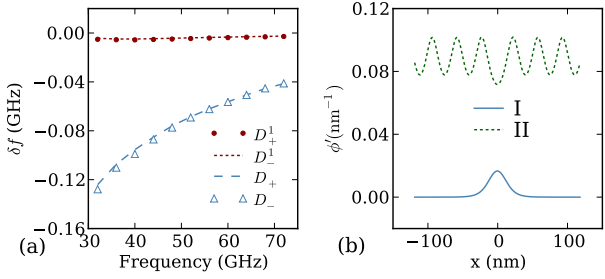

FIG. S3. (a) Plot of the frequency difference $\delta f$ versus frequency. The used anisotropies are $K_\perp = 2 \times 10^5$ J/m$^3$ and $K = 1 \times 10^5$ J/m$^3$. DMI constants are $D_\pm^1 = \pm 4 \times 10^{-4}$ J/m$^2$ and $D_\pm = \pm 1.58 \times 10^{-3}$ J/m$^2$. (b) Two types of domain walls obtained from simulation data of type I with $K_\perp = 8 \times 10^5$ J/m$^3$ and type II with $K_\perp = 8 \times 10^4$ J/m$^3$.

through the domain wall, which is true for $D = 0$ or $K_\perp = 0$ case. However, this assumption is not strictly true in the scenario $D \neq 0$ and $K_\perp > 0$. Fig. S3(a) shows the frequency difference $\delta f$ as a function of frequency with different DMI constants. It is found that the frequency differences are negative for both positive and negative DMI constants. Note that for relatively weak DMI case such as $|D^1| = 4 \times 10^{-4}$ J/m$^2$ the frequency difference $\delta f$ is not significant.

## THE ESTIMATION OF THE DOMAIN WALL WIDTH

In the main content we discussed the domain wall profile for the $K_\perp = 0$ case. By assuming the constant DW width $\Delta$ and the linear dependence of $\phi$ we obtain $\Delta = \sqrt{A/(K - A/\xi^2)}$ where $\xi = 2A/D$ is the characteristic length. However, these assumptions are not true for $K_\perp > 0$ case. To estimate the DW width in general case we start with the differential equations

$$2A\theta'' = \sin 2\theta (A\phi'^2 + K + K_\perp \sin^2\phi - D\phi'),$$
$$\sin\theta(2A\phi'' - K_\perp \sin 2\phi) = 2\cos\theta(D - 2A\phi')\theta'. \quad (S4)$$

We assume that the DW profile can be described by,

$$\cos\theta = -\tanh f(x),$$
$$\sin\theta = \text{sech}\, f(x), \quad (S5)$$

where $f(x)$ is an unknown function of space and thus the DW width is related to $f'(x)$. We expect $f(x)$ is a monotonic non-decreasing function. From Eq. (S5) we know that $\theta' = \sin\theta f'(x)$ and $\theta'' = \sin\theta[\cos\theta f'(x)^2 + f''(x)]$, and by considering the first equation of Eq. (S4) we assume $f''(x) = g(x)\cos\theta$ since $f''(x) = 0$ leads to a constant DW width. Therefore, we arrive at,

$$A(f'^2 + g) = A\phi'^2 + K + K_\perp \sin^2\phi - D\phi' \quad (S6)$$

Note that for the $K_\perp = 0$ case $f'^2 + \phi'^2 = K/A$ and by combining it with the second equation of Eq. (S4), we assume

$$A(f'^2 + \phi'^2) = K + K_\perp \sin^2\phi, \quad (S7)$$

which leads to $Ag = (2A\phi' - D)\phi'$ and further more,

$$Af'' = -(2A\phi' - D)\phi'\tanh f. \quad (S8)$$

We can check that Eq. (S7) and Eq. (S8) are the solutions of Eq. (S4). The asymptotic behavior of $f''$ can be obtained from Eq. (S8) by making use of the fact that $\tanh f \to 1$ in the limit of $x \to +\infty$. We can distinguish two types of domain walls according to the value of $f''_\infty = \lim_{x \to +\infty} f''(x)$.

The first case is that $f''_\infty = 0$, which requires $\phi'_\infty = D/(2A)$ or $\phi'_\infty = 0$. Note that $\phi'_\infty = D/(2A)$ corresponds to the DW profile obtained in the main content, which is only valid for $K_\perp = 0$ case. Therefore, we identify the first type of the DW with $\phi'_\infty = 0$. As an approximation, we assume $\phi = a\tanh(\nu x)$ for this type of wall, so we have $f'_0 = \sqrt{K/A - a^2\nu^2}$ and $f'_\infty = \sqrt{(K + K_\perp \sin^2 a)/A}$. As we can see, the DW width is bounded by $f'_0$ and $f'_\infty$. By using $\nu \sim \sqrt{K/A}$ we find that $a \to 0$ in the limit of $\kappa \to \infty$ and thus $\Delta \to \sqrt{A/K}$.

The second case is that $f''_\infty$ does not converge and $\phi(x)$ is a monotonic function. From Eq. (S7) one can deduce that at least one of $f'$ and $\phi'$ oscillates, and combining Eq. (S8) we conclude that both of them oscillate. In this scenario, we take the following approximation forms for $f \gg 1$,

$$\phi' = \eta - c\cos(2\phi - \phi_0),$$
$$f' = \lambda + c\sin(2\phi - \phi_0), \quad (S9)$$

where $\eta = 1/\xi = D/(2A)$. We can find that this assumption satisfies Eq. (S8). Substitute Eq. (S9) into Eq. (S7) and to eliminate the term $\sin 2\phi$ one obtains $\sin\phi_0 = \lambda/\sqrt{\lambda^2 + \eta^2}$ and $\cos\phi_0 = \eta/\sqrt{\lambda^2 + \eta^2}$. Finally, we arrive at,

$$\lambda^2 = \frac{K}{A}\frac{1 + \kappa/2 + \sqrt{1+\kappa}}{2} - \eta^2, \quad (S10)$$

and thus the domain wall width $\Delta_\infty = 1/\lambda$ can be established. The oscillation amplitude is given by $c = K_\perp/(4A\sqrt{\eta^2 + \lambda^2})$. This solution is only valid for weak easy plane anisotropy case that requires $c < \eta$. And thus the critical $\kappa_c$ can be obtained by solving the equation $K\kappa_c = 2D\sqrt{\eta^2 + \lambda^2}$, i.e., $AK\kappa_c^2 = 2D^2(1 + \kappa_c/2 + \sqrt{1+\kappa_c})$. Fig. S3(b) shows the two types of domain walls extracted from simulation data.

# THE INFLUENCE OF THE MAGNETIZATION TILTING AT EDGES

In the main article we considered a 1d model in which the magnetization tilting is absent. In this section we perform the simulations in a 2d sample. Fig. S4(a) shows

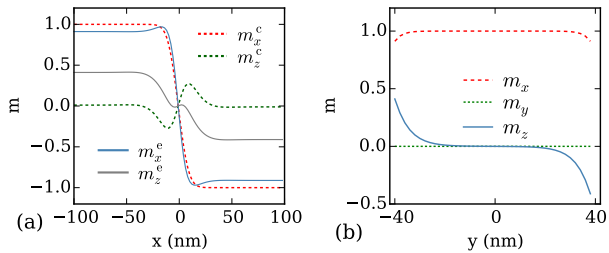

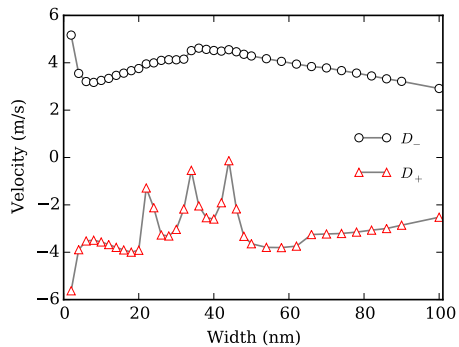

FIG. S4. (a) The domain wall profile shown through the magnetization components $m_x^c$ and $m_z^c$ at the center ($y = 0$) and the components $m_x^e$ and $m_z^e$ at the edge ($y = -40$ nm) along the $x$-axis. (b) A cross section of the three magnetization components across the width of the 2d sample for the domains (at $x = -500$ nm).

the domain wall profile along the center (magnetization components $m_z^c$ and $m_x^c$ for $y = 0$) and the edge ($m_x^e$ and $m_z^e$ for $y = -40$ nm) of the sample. The width of the 2d sample in $y$-direction is 80 nm (going from $-40$ nm to $+40$ nm) and the cell size we have used is $2 \times 2$ nm$^2$. The domain wall profile is obtained with the anisotropies $K_\perp = 2 \times 10^5$ J/m$^3$ and $K = 1 \times 10^5$ J/m$^3$. It is found that the magnetization component $m_z$ at the edge is significant ($|m_z| \sim 0.4$) compared to the $m_z$ ($\sim 0$) in the center due to the DMI. The magnetization tilting decreases to zero rapidly when moving away from the edges, as shown in Fig. S4(b), and magnetization component $m_y$ is not tilted in this 2d sample for the head-to-head or tail-to-tail domain wall profile.

FIG. S5. The domain wall velocity as a function of the sample width for different DMI constants with the anisotropies $K_\perp = 2 \times 10^5$ J/m$^3$ and $K = 1 \times 10^5$ J/m$^3$. The frequency of the ac field is $f = 28$ GHz.

We fix the sample length $L_x = 2000$ nm and change the film width to probe the influence of this magnetization tilting on the domain wall dynamics. Fig. S5 shows the influence of the sample width as well as the sign of the DMI constant on the domain wall velocity. It can be seen that the domain wall in the narrowest sample (first data point for width of 2 nm) has the highest velocity, corresponding to the 1d case. Moving to the next data point (4 nm width), the magnitude of the domain wall velocity decays by about a third, both for the positive $D_+$ and negative $D_-$ values of $D$. For the widths greater than approximately 60 nm, the magnitude of the domain wall velocity decays slowly as a function of increasing width. For geometry widths between 4 and 60 nm, the velocity functions show non-trivial behaviour, and the curve $D_+$ for positive $D$ exhibits peaks at particular width values. The details of these features justify a study in its own right and are outside the scope of this work. We summarize, that in a 2d-system, the domain wall velocity is somewhat smaller than in the 1d-system, but remains of the same order of magnitude.



\end{document}

%% file: paper.bbl
%